\documentclass[a4paper]{jpconf}
\usepackage{graphicx}
\usepackage{amsfonts}
\usepackage{amssymb}
\usepackage{amsmath}
\usepackage{marvosym}
\usepackage{enumitem}
\usepackage{cite}
\begin{document}

\title{Neutrino Mass Sum Rules}

\author{Martin Spinrath}

\address{Institut f\"ur Theoretische Teilchenphysik, Karlsruhe Institute of Technology, Engesserstra\ss{}e 7, D-76131 Karlsruhe, Germany}

\ead{martin.spinrath@kit.edu}

\footnotesize

\begin{abstract}
Neutrino mass sum rules are an important class of predictions in flavour models relating the
Majorana phases to the neutrino masses. This leads, for instance, to enormous restrictions
on the effective mass as probed in experiments on neutrinoless double beta decay. While
up to now these sum rules have in practically all cases been taken to hold exactly, we will go
here beyond that. 
While the effect of the renormalisation group running can be visible, the qualitative
features do not change. This changes somewhat for model dependent corrections which might
alter even the qualitative predictions but only for large corrections and a high neutrino mass scale
close to the edge of the current limits.
This finding backs up the solidity of the predictions derived in
the literature apart from some exceptions, and it thus marks a very important step in deriving
testable and robust predictions from neutrino flavour models.
\end{abstract}

\section{Setup}

In \cite{King:2013psa,Gehrlein:2015ena,Gehrlein:2016wlc} a general
parametrisation for neutrino mass sum rules was given
\begin{equation}
\label{eq:parametrisation_SR}
s \equiv c_1 \left(m_1 \text{e}^{-\text{i}\,\phi_{1}}\right)^d 
\text{e}^{\text{i}\Delta \chi_{13}}+
c_2 \left(m_2 \text{e}^{-\text{i}\,\phi_{2}}\right)^d 
\text{e}^{\text{i}\Delta \chi_{23}}
+m_3^d~ \stackrel{!}{=} 0 \;,
\end{equation}
where $m_i$ are the light neutrino masses and $\phi_i$ the Majorana phases.
All the other coefficients are given by one of the twelve known sum rules,
cf.~table~\ref{tab:sumrules}. For a perturbed sum rule in \cite{Gehrlein:2016wlc} 
we also defined a normalised sum rule, $\hat{s}$, and the first order perturbation
of it, $\delta \hat{s}$, by
\begin{equation}
 \hat{s} \equiv \frac{ s}{m_n^d} \text{ and } \delta \hat{s} \equiv \frac{\delta s}{m_n^d} \;, \label{eq:dsh}~
\end{equation}
where $m_n$ is chosen in such a way that there is no artificial enhancement by
a factor $m_i/m_j \gg 1$.

\begin{table}
\caption{\footnotesize
 Summary table of the sum rules taken from \cite{Gehrlein:2015ena,Gehrlein:2016wlc},
 cf.~Eq.~\eqref{eq:parametrisation_SR}. 
 \label{tab:sumrules}
} 
\footnotesize\rm
\begin{center}
\begin{tabular}{c c c c c c c}
\br
Sum rule & References & $c_1$&$c_2$&$d$&$\Delta \chi_{13}$&$\Delta \chi_{23}$ \\
\mr
    1& \cite{Barry:2010yk,Bazzocchi:2009da,Ding:2010pc,Ma:2005sha,Ma:2006wm,Honda:2008rs,Brahmachari:2008fn,Kang:2015xfa,Everett:2008et, Boucenna:2012qb} &$1$&$1$&$1$&$\pi$&$\pi$\\
    2& \cite{Mohapatra:2012tb} &$1$&$2$&$1$&$\pi$&$\pi$\\
    3& \cite{Barry:2010yk,Ma:2005sha,Ma:2006wm,Honda:2008rs,Brahmachari:2008fn,Altarelli:2005yx,Chen:2009um,Chen:2009gy,Kang:2015xfa,Altarelli:2005yp, Altarelli:2006kg, Ma:2006vq, Bazzocchi:2007na, Bazzocchi:2007au, Lin:2008aj, Ma:2009wi, Ciafaloni:2009qs, Bazzocchi:2008ej, Feruglio:2013hia, Chen:2007afa, Ding:2008rj, Chen:2009gf, Feruglio:2007uu, Merlo:2011hw, Luhn:2012bc, Fukuyama:2010mz} &$1$&$2$&$1$&$\pi$&$0$\\
    4& \cite{Ding:2013eca, Lindner:2010wr} &$1/2$&$1/2$&$1$&$\pi$&$\pi$\\
    5& \cite{Hashimoto:2011tn} &$\tfrac{2}{\sqrt{3}+1}$&$\tfrac{\sqrt{3}-1}{\sqrt{3}+1}$&$1$&$0$&$\pi$\\
    6& \cite{Barry:2010yk,Bazzocchi:2009da,Ding:2010pc,Cooper:2012bd,Ding:2011cm,Gehrlein:2014wda } &$1$&$1$&$-1$&$\pi$&$\pi$\\
    7& \cite{Barry:2010yk,Altarelli:2005yx,Chen:2009um,Chen:2009gy,Altarelli:2009kr,Altarelli:2008bg, Morisi:2007ft, Adhikary:2008au, Lin:2009bw, Csaki:2008qq, Hagedorn:2009jy, Burrows:2009pi, Ding:2009gh, Mitra:2009jj, delAguila:2010vg, Burrows:2010wz, Ahn:2014zja, Karmakar:2014dva, Ahn:2014gva} &$1$&$2$&$-1$&$\pi$&$0$\\
    8& \cite{He:2006dk, Berger:2009tt, Kadosh:2010rm, Lavoura:2012cv} &$1$&$2$&$-1$&$0$&$\pi$\\
    9& \cite{King:2012in} &$1$&$2$&$-1$&$\pi$&$\pi/2,3\pi/2$\\
    10& \cite{Adulpravitchai:2009re,Hirsch:2008rp} &$1$&$2$&$1/2$&$\pi,0,\pi/2$ &$0,\pi,\pi/2$ \\
    11& \cite{Dorame:2011eb} &$1/3$&$1$&$1/2$&$\pi$&$0$\\
    12& \cite{Dorame:2012zv} &$1/2$&$1/2$&$-1/2$&$\pi$&$\pi$\\
\br
\end{tabular}
\end{center}
\end{table}

\section{Selected Results}

Some sum rules allow only for one of the two possible neutrino mass orderings.
This can be most easily understood from a geometrical interpretation of the sum
rule as a closed triangle in a complex plane \cite{King:2013psa}. In sum rule~2,
for instance, one can then easily show that inverted ordering is excluded \cite{Gehrlein:2015ena} because
\begin{equation}
\text{cos}\, \alpha^{\text{tree}} = \frac{m_1^2 - 4 m_2^2 - m_3^2}{4 m_2 m_3} < - \frac{1}{4} \left( 3 \frac{m_2^2}{m_3^2} + 1 \right) < -1 \Rightarrow \text{\Large \Lightning} \;,
\end{equation}
where $\alpha$ is one of the angles in the triangle.
The RGE corrections to $\cos \alpha$  in the MSSM \cite{Gehrlein:2015ena},
\begin{equation}
\delta (\text{cos}\, \alpha)^{\text{RGE}} \approx - \underbrace{\frac{C y_\tau^2}{192 \pi^2}}_{>0} \underbrace{\frac{2.8 m_1^2 - 0.4 m_2^2 + 0.1 m_3^2}{m_2 m_3}}_{>0} \underbrace{\text{log}\, \frac{M_S}{M_Z}}_{>0} < 0 \;,
\end{equation}
make things worse. This statement is true for most of the sum rules in the
overwhelming part of the parameter space. For the very few cases where the
sign is correct one would still need very extreme parameter choices, 
which are already disfavoured, see also \cite{Spinrath:2016cwn}.

For general corrections forbidden orderings can be reconstituted \cite{Gehrlein:2016wlc}.
For sum rule 2 forbidden ordering could be reconstituted by a 30\% correction
and a neutrino mass scale of 0.05~eV, which is on the edge of being disfavoured by
cosmology. For smaller mass scales the necessary corrections grow quickly, see
Table~3 in \cite{Gehrlein:2016wlc} and indeed in most cases (except for sum rule 10) the
newly allowed regions are not very big and practically excluded by the cosmological bounds on the neutrino mass scale.

\section{Summary}

Neutrino mass sum rules are a very powerful tool to test and
discriminate flavour models \cite{King:2013psa, Agostini:2015dna}. This was confirmed in the systematic
studies of RG corrections \cite{Gehrlein:2015ena} and general
perturbations \cite{Gehrlein:2016wlc} where it was shown that
many essential predictions still valid at least qualitatively.

In the future experimental information on the neutrino mass scale,
their mass ordering, neutrinoless double beta decay, but also
information about new physics at the TeV scale like supersymmetry
will be a crucial test to understand if a neutrino mass sum rule
is realised in nature.

\ack

MS is  supported  by  BMBF  under  contract no.\ 05H12VKF and
acknowledges partial support to attend the NEUTRINO 2016 conference
by Deutscher Akademischer Auslandsdienst (DAAD).

\section*{References}
\bibliography{Neutrino_Spinrath_lit}

\end{document}